\documentclass[]{interact}
\usepackage{amssymb,amsmath,amsfonts,mathrsfs}

\usepackage{epstopdf}
\usepackage{subfigure}

\usepackage{natbib}
\usepackage[font=large, labelsep=space]{caption}

\bibpunct[, ]{(}{)}{;}{a}{}{,}

\theoremstyle{plain}
\newtheorem{theorem}{Theorem}[section]

\newtheorem{corollary}[theorem]{Corollary}
\newtheorem{proposition}[theorem]{Proposition}

\theoremstyle{definition}

\theoremstyle{remark}

\begin{document}


\title{ Design and performance evaluation in  Kiefer-Weiss problems when sampling from discrete exponential families}

\author{
\name{Andrey Novikov\textsuperscript{}\thanks{CONTACT A. Novikov, Universidad Aut\'onoma  Metropolitana - Unidad Iztapalapa, Avenida Ferrocarril San Rafael Atlixco 186, col. Leyes de Reforma 1A Secci\'on, C.P. 09310, Cd. de  M\'exico, M\'exico. Email: an@xanum.uam.mx} and Fahil Farkhshatov\textsuperscript{}}
\affil{\textsuperscript{} Metropolitan Autonomous University, Mexico City, Mexico}
}
\maketitle
\begin{abstract}
In this paper, we deal with problems of testing hypotheses in the framework of sequential statistical analysis. 
The main concern is the optimal design and performance evaluation of sampling plans in the Kiefer-Weiss problems.

 For the observations which follow a discrete exponential family, we provide algorithms for
optimal design in the  modified Kiefer-Weiss problem, and obtain formulas for evaluating their performance, calculating operating characteristic function, average sample number, and some related characteristics. These formulas cover, as a particular case, the sequential probability ratio tests (SPRT) and their truncated versions, as well as optimal finite-horizon sequential tests. 

On the basis of the developed algorithms we propose a method of construction of  optimal tests and their performance evaluation for the original  Kiefer-Weiss problem.

All the algorithms are implemented as functions in the R programming  language and can be 
downloaded from
{\tt https://github.com/tosinabase/Kiefer-Weiss}, where the functions for the binomial, Poisson, and  negative binomial distributions are readily available. 

Finally, we make numerical comparisons of the Kiefer-Weiss solution with the SPRT and the fixed-sample-size test having the same levels of the error probabilities.
\end{abstract}

\begin{amscode}
 62L10, 62L15, 62F03, 60G40, 62M02
 \end{amscode}

\section{Introduction}
%
%
%

In a sequential statistical  experiment, a sequence of random variables $X_1,X_2$, $\dots, X_n, \dots$ is potentially  available to the statistician on the one-by-one basis. The observed data bear the  information  about  the underlying distribution $P_\theta$,  $\theta$  being an unknown parameter whose true value  is of interest to the statistician. In this paper, we are concerned with testing  a simple hypothesis H$_0:$ $\theta={\theta_0}$ against a simple alternative H$_1:$ $\theta={\theta_1}$, in the framework of sequential analysis, more specifically, 
in  the Kiefer-Weiss setting \citep[see][]{Kiefer}.

In its simplest form,
a sequential hypothesis test is a pair $\langle \tau, \delta\rangle$ consisting of a stopping time  $\tau$  and a (terminal) decision rule $\delta$. 
Formally, it is required that $\{\tau= n\}\in \sigma(X_1,\dots, X_n)$ and $\{\tau=n,\delta=i\}\in\sigma(X_1,\dots, X_n)$, for any natural $n$ and for $i=0,1$.
The performance characteristics of  a sequential test are the type I and type II error probabilities,
$\alpha(\tau,\delta)=P_{\theta_0}(\delta=1)$ and $\beta(\tau,\delta)=P_{\theta_1}(\delta=0)$,
and the average sample number, $E_\theta \tau$.

The Kiefer-Weiss problem consists in  finding a test $\langle\tau,\delta\rangle$ with a minimum value  of 
$\sup_\theta E_\theta \tau$, 
among all the tests satisfying the constraints on the type I and type II error probabilities:  
\begin{equation*}\label{2}
\alpha(\tau,\delta)\leq\alpha\quad\mbox{and}\quad\beta(\tau,\delta)\leq\beta.
\end{equation*}

This article is meant to be a continuation of our recent publication \cite{novikov2021comp} on the Kiefer-Weiss problem, where we investigated the case of sampling from a Bernoulli population. 
A brief  review of relevant general results concerning the Kiefer-Weiss problem
can be found in the Introduction to
\cite{novikov2021comp}.

We consider in this paper  models of i.i.d. observations that follow a discrete exponential (Koopman-Darmois) family. 
We are concerned with the Kiefer-Weiss and related problems, where
 we propose a unified approach to the optimal design and performance evaluation of sampling plans. We develop the whole set of computational  algorithms for this. The algorithms are fully implemented  in the   R programming language \citep{R} for the binomial, Poisson and negative binomial (Pascal) distributions. Performance evaluation for SPRT and truncated SPRT is implemented as well. 

The program code can be downloaded from the GitHub repository at {\tt https://github.com/tosinabase/Kiefer-Weiss}. 

 Using the developed computer code, we calculate in the implemented models, for a range of error probabilities $\alpha, \beta$, 
 the parameters of optimal sampling plans and their characteristics, as well as those of the Wald's sequential probability ratio tests (SPRT) and the fixed-sample-size tests (FSST) with the same levels of type I and type II error probabilities.
 
In  Section 2, we lay out  general results our solutions to the Kiefer-Weiss problem are based on, for the models we are concerned with.

In Section 3, we develop computing algorithms 
for optimal design and performance evaluation of the respective sequential sampling plans and present the numerical results.

Section 4 contains a discussion of the results and marks some direction for further work.
 
\section{ Kiefer-Weiss problem}

In this section, we formulate  general results on the structure of the optimal tests in the Kiefer-Weiss problem and its modified version.

\subsection{Optimal sampling plans}

We follow in this section the notation and general assumptions of \cite{Novikov2009}.

We consider here non-randomized sequential tests $\langle\psi,\phi\rangle$, with  $\psi=(\psi_1,\psi_2,$ $\dots)$ being a stopping rule, and $\phi=(\phi_1,\phi_2,\dots)$ being a (terminal) decision rule. It is  assumed that $\psi_n=\psi_n(x_1,\dots,x_n)$ and $\phi_n=\phi_n(x_1,\dots,x_n)$ are measurable indicator functions. After the data $x_1, \dots,x_n$  have been observed, $\psi_n(x_1,\dots,x_n)=1$ means stopping at stage $n$,
and $\phi(x_1,\dots,x_n)=i$ means accepting hypothesis H$_i$ after the decision to stop has been made, $i =0$ or $1$, for any $n=1,2,\dots$ Let us denote $\tau_\psi$ the stopping time generated by the stopping rule $\psi$.

Denote $c_n^\psi=c_n^\psi(x_1,\dots,x_n)=(1-\psi_1)(1-\psi_2)\dots(1-\psi_{n-1})$, and $s_n^\psi=c_n^\psi\psi_n$, so for any test 
\begin{equation*}
\alpha(\psi,\phi)=\sum_{n=1}^{\infty}E_{\theta_0}s_n^\psi\phi_n
\end{equation*}
is the type I error probability,  \begin{equation*}
\beta(\psi,\phi)=\sum_{n=1}^{
\infty}E_{\theta_1}s_n^\psi(1-\phi_n)
\end{equation*}
is the type II error probability, and
\begin{equation*}
N(\theta;\psi)=\sum_{n=1}^{\infty}nE_{\theta}s_n^\psi=E_\theta \tau_\psi
\end{equation*}
is the average sample number when the true parameter value is $\theta$
(provided that $\sum_{n=1}^{\infty}E_{\theta}s_n^\psi(=P_\theta(\tau_\psi<\infty))=1$, -- otherwise it is infinite).

It is  common to express the error probabilities in terms of the operating characteristic curve defined as
 \begin{equation*}
OC_\theta(\psi,\phi)=\sum_{n=1}^{
\infty}E_{\theta}s_n^\psi(1-\phi_n):
\end {equation*}
$\alpha(\psi,\phi)=1-OC_{\theta_0}(\psi,\phi),\quad\beta(\psi,\phi)=OC_{\theta_1}(\psi,\phi)
$.

Let $\mathscr S(\alpha,\beta)$ be the set of all tests such that
\begin{equation}\label{7_1}
\alpha(\psi,\phi)\leq \alpha,\quad \beta(\psi,\phi)\leq \beta,
\end{equation}
where $\alpha,\beta\in[0,1]$ are some fixed real numbers. 

We are interested in finding  tests  minimizing $\sup_\theta N(\theta;\psi)$ over all the tests  in $\mathscr S(\alpha,\beta)$.
This problem is known as the Kiefer-Weiss problem. 

The respective modified Kiefer-Weiss problem is to minimize $N(\theta_{};\psi)$ over all $\langle \psi,\phi\rangle\in \mathscr S(\alpha,\beta)$, for a given fixed value of $\theta_{}$.

All the known solutions of the Kiefer-Weiss problem, for particular models, have been obtained through the modified version of it \citep[see][among many others]{Kiefer,Weiss,FreemanWeiss,Lai1973,Lorden,Huffman1983,Zhitlukhin, Tartakovsky2014}.

The solutions to the modified Kiefer-Weiss problem can be obtained, at least in theory,
in a very general situation using the following variant of the Lagrange multipliers method. 

Let \begin{equation}\label{10a_1}L_\theta(\psi,\phi)=N(\theta_{};\psi)+\lambda_0\alpha(\psi,\phi)+\lambda_1\beta(\psi,\phi),\end{equation} where $\theta_{} $ is some fixed value of the parameter and $\lambda_0,\lambda_1$ are some nonnegative constants (called Lagrange multipliers).

Then the tests minimizing $N(\theta_{};\psi)$ subject to  \eqref{7_1} can be obtained 
through an unconstrained minimization of $L_\theta(\psi,\phi)$ over all $\langle\psi,\phi\rangle$,
 using  an appropriate choice of the Lagrange multipliers  \citep[see][Section 2]{Novikov2009}.

  \cite{Lorden} shows that in the case of i.i.d. observations the problem of minimizing the Lagrangian function is reduced to an optimal stopping problem for a Markov process.

  It is easy to see that finding Bayesian tests used in \cite{Kiefer} is mathematically equivalent to the minimization of \eqref{10a_1}.

  To construct the optimal tests we need an additional assumption on the distribution of the observations. Let $f_\theta^n=f_\theta^n(x_1 ,\dots, x_n)$ be the Radon--Nikodym derivative of the distribution of $X_1,\dots,X_n$ with respect to a product-measure $\mu^n=\mu\otimes\dots\otimes \mu$ ($n$ times $\mu$ by itself), $n=1,2,\dots$.

  In \cite{Lorden}, it is shown that, in the case of i.i.d. observations following  a distribution from a Koopman-Darmois family, the tests giving solution to the modified problem have  bounded with probability one stopping times when $\theta_{}\in(\theta_0,\theta_1)$.

Let us describe the construction of tests minimizing the Lagrangian function calculated at some $\theta_{}$,  over all {\em truncated} tests, i.e. those not taking more than a fixed number $H$ of observations ($H$ is also called {\em horizon} in this case).

Formally, let
$\mathscr S^H=\{\langle\psi,\phi\rangle:   \; c_{H+1}^\psi\equiv 0 \}$
be the class of all such tests.

The structure of tests minimizing the Lagrangian function in $\mathscr S^H$
can be characterized in the following way.

Let us define \begin{equation}\label{14a_1}V_{\theta,H}^H=\min\{\lambda_0f_{\theta_0}^H,\lambda_1f_{\theta_1}^H\},\end{equation} and, recursively over $n=H-1,\dots,1$,
\begin{equation}\label{11a_1}
V_{\theta,n}^H=\min\left\{\lambda_0f_{\theta_0}^{n},\lambda_1f_{\theta_1}^{n},
f_{\theta_{}}^{n}+\mathcal{I}V^H_{\theta,n+1} 
\right\},
\end{equation}
being
\begin{equation}\label{14a_2}
\mathcal{I}V^H_{\theta,n+1}=\left(\mathcal{I}V^H_{\theta,n+1}\right)(x_1,\dots,x_{n})=\int V^H_{\theta,n+1}(x_1,\dots,x_{n+1})d\mu(x_{n+1}).
\end{equation}
Note that functions $V_{\theta,n}^H$ defined above, as well as $L_\theta(\psi,\phi)$ in \eqref{10a_1}, implicitly depend on $\lambda_0,\lambda_1$. 

From the Proposition 1 of \cite{novikov2021comp} we obtain that 
for any $\langle \psi,\phi\rangle\in \mathscr S^H$
\begin{equation}\label{11a_2}
L_\theta(\psi,\phi)\geq 1+\mathfrak I V_{\theta,1}^H,
\end{equation}
and the right-hand side in \eqref{11a_2} is attained if 
\begin{equation}\label{11a_3}
\psi_n= I_{
\{ \min\{\lambda_0f_{\theta_0}^{n},\lambda_1f_{\theta_1}^{n}\}\leq
f_{\theta_{}}^{n}+\mathcal{I}V^H_{\theta,n+1}       \}}\end{equation} 
$\mbox{for}\; 
n=1,2,\dots, H-1,\;\mbox{and}\; \psi_H\equiv 1$, and
\begin{equation}\label{11a_4}
\phi_n=I_{\{
\lambda_0f_{\theta_0}^n\leq\lambda_1f_{\theta_1}^n
\}}
\end{equation}
for $n=1,2,\dots, H$.  Any test with strict inequalities in \eqref{11a_3} and/or \eqref{11a_4} for some (or all) $n$ also attains the right-hand side in \eqref{11a_2}.
Let us denote $\mathscr M^H_\theta$ the set of all such tests, and let $\mathscr M^H=\cup_\theta\mathscr M^H_\theta$.

Under very mild conditions, the optimal non-truncated tests are obtained on the basis of limits $V_{\theta,n}=\lim_{H\to\infty}V_{\theta,n}^H$. Optimal stopping rules for the non-truncated tests are obtained substituting $V_{\theta,n}$ for $V_{\theta,n}^H$ in \eqref{11a_3}: 
\begin{equation}\label{4n_1}
\psi_n= I_{
\{ \min\{\lambda_0f_{\theta_0}^{n},\lambda_1f_{\theta_1}^{n}\}\leq
f_{\theta_{}}^{n}+\mathcal{I}V_{\theta, n+1}       \}}\end{equation}  for all natural $n$, 
 applying \eqref{11a_4} for all $n$ as the terminal decision rule (see Section 3 in \cite{Novikov2009}). When the optimal test is truncated, it holds $V_{\theta,n}=V_{\theta,n}^H$ for large enough $H$.
Let $\mathscr M_\theta$ be the class of tests $\langle\psi,\phi\rangle$ satisfying \eqref{11a_4} and \eqref{4n_1}  
 for all $n$ (including those with strict inequalities), and $\mathscr M=\cup_\theta\mathscr M_\theta$. 

Proposition 2 in \cite{novikov2021comp} offers the following way of solving the Kiefer-Weiss problem.

Take some $\langle\psi^*,\phi^*\rangle\in \mathscr M_\theta$, and let \begin{equation} \label{20m_1}\sup_\vartheta N(\vartheta;\psi^*)-N(\theta;\psi^*)=
\Delta. \end{equation}
Then
\begin{equation}\label{4n_2}
\sup_\vartheta N(\vartheta;\psi^*)\leq\inf \sup_\vartheta N(\vartheta;\psi)+\Delta,\end{equation} 
where the infimum is taken over all tests $\langle\psi,\phi\rangle$ such that $\alpha(\psi,\phi)\leq\alpha(\psi^*,\phi^*)$ and $\beta(\psi,\phi)\leq\beta(\psi^*,\phi^*)$. It follows from \eqref{4n_2} that $\langle\psi^*,\phi^*\rangle$ is as close as to within $\Delta$ to the solution to the Kiefer-Weiss problem. In symmetrical cases, $\Delta=0$ if one takes $\theta=(\theta_0+\theta_1)/2$ \citep[see][among others]{Kiefer, Weiss, Lai1973, Lorden}.

 In \cite{novikov2021comp} we proposed a numerical solution seeking for a minimum value of $\Delta $, over $\mathscr M$, by means of a computer program. For the case of Bernoulli observations, we wrote a program code in R language for the numerical minimization, and ran it for a broad range of $\theta_0,\theta_1$ and $\alpha$ and $\beta$. In all evaluated examples the minimum value of $\Delta$ was very close to 0 \citep[see][]{novikov2021comp}. 
Below in this paper, we show that the same happens when sampling from binomial, Poisson and geometric distributions.

\section{Algorithms for computing optimal sampling plans   }

In this section, we describe computational algorithms for the modified Kiefer-Weiss problem in the case when the hypothesized distributions come from a discrete  exponential family. They are based on the algorithms we developed for sampling from a Bernoulli population \citep[see][]{novikov2021comp}. The application to the Kiefer-Weiss problem straightforwardly follows  from    Proposition 2 therein, in fact,  the algorithmic part stays unchanged. 

The proposed algorithms are implemented in the form of program code in the R programming language \citep{R} and are available as a part of GitHub repository at {\tt https://github.com/tosinabase/Kiefer-Weiss}.

\subsection{Optimal tests in modified Kiefer-Weiss problem}

Let the observations $X_1,X_2,\dots,X_n,\dots$ be independent identically distributed  random variables following a distribution from a  discrete one-parametric exponential family. More specifically, we assume that

\begin{equation}\label{9d_1}f_\theta(x)=\exp(\theta x-b(\theta))h(x), \;x\in\mathbb Z^+,\end{equation} 
 where $h(x)$ is a non-negative function 
on $\mathbb Z^+$. 
Then the joint probability \begin{equation*}f_\theta^n(x_1,\dots,x_n)=\exp(\theta s_n-nb(\theta))h_n(x_1,\dots,x_n),\end{equation*} where $s_n=\sum_{i=1}^n x_i$, and $h_n(x_1,\dots,x_n)=\prod_{i=1}^nh(x_i)$,
$n=1,2,3, \dots$.

Let  $\theta_0<\theta_1$ be two fixed  parameter values defining the hypotheses H$_0$ and H$_1$.

Let us construct a solution to the modified Kiefer-Weiss problem for a given $\theta_{}$ when sampling from a distribution of type \eqref{9d_1}.

Let us express \eqref{14a_1} -- \eqref{14a_2} in a more ``computer-friendly'' form, namely,  using the distribution of the sufficient statistic $\sum_{i=1}^n X_i$ \citep[see][]{novikov2021comp}. This helps to avoid dealing with very small numbers representing the joint probabilities in \eqref{14a_1}, etc., in case the truncation level $H$ is high.

Let
\begin{equation*}\label{20d_3}
g_\theta^n(s)=C_n(s)\exp(\theta s-nb(\theta)),\; s\in \mathbb Z^+,
\end{equation*}
be the probability mass function of the statistic $\sum_{i=1}^n X_i$, for any natural $n$.

Now, let us define
\begin{equation*}\label{7n_1}U_{\theta,H}^H(s)=\min\{\lambda_0g_{\theta_0}^H(s),\lambda_1g_{\theta_1}^H(s)\},\;s\in\mathbb Z^+, \end{equation*}
and, recursively over $n=H,\dots, 2$,
\begin{equation*}\label{7n_2}
U_{\theta,n-1}^H(s)=\min\{\lambda_0g_{\theta_0}^{n-1}(s),\lambda_1g_{\theta_1}^{n-1}(s),
g_{\theta}^{n-1}+\mathcal J_n U^H_{\theta,n}(s)\},\;s\in\mathbb Z^+,\end{equation*}
where $\mathcal J_n U(s)=\sum_xU(s+x)d_n(x,s),\;s\in\mathbb Z^+,$
 being \begin{equation*}\label{20d_4}d_n(x,s)=C_1(x)\frac{C_{n-1}(s)}{C_{n}(s+x)}, \;x,s\in\mathbb{Z}^+.\end{equation*} 

\begin{proposition}\label{P1}
\begin{equation}\label{8n_11}V_{\theta,n}^H(x_1,\dots,x_n)=U_{\theta,n}^H(s_n)h_n(x_1,\dots,x_n)/C_n(s_n),\end{equation} where  $s_n=\sum_{i=1}^nx_i$, for  $n=1,\dots, H$. 
\end{proposition}
Proof. By induction over $n=H,H-1,\dots, 1$.

It follows from \eqref{14a_1} that
\begin{eqnarray*}V_{\theta,H}^H(x_1,\dots,x_H)=\min\{\lambda_0f_{\theta_0}^H(x_1,\dots,x_H),\lambda_1f_{\theta_1}^H(x_1,\dots,x_H)\}\\
=\min\{\lambda_0g_{\theta_0}^H(s_H),\lambda_1g_{\theta_1}^H(s_H)\}\frac{h_H(x_1,\dots,x_H)}{C_H(s_H)}\\
=U_{\theta,H}^H(s_H)\frac{h_H(x_1,\dots,x_H)}{C_H(s_H)}.
\end{eqnarray*}

Let us suppose that \eqref{8n_11} holds for some $2\leq n\leq H$. Then,
by virtue of \eqref{11a_1}
\begin{eqnarray*}\label{8n_12}
V_{\theta,n-1}^H=\min\left\{
\lambda_0f_{\theta_0}^{n-1},\lambda_1f_{\theta_1}^{n-1},f_{\theta_{}}^{n-1}
+\sum_{x_n}U^H_{\theta,n} (x_1+\dots+x_n)\frac{h_n(x_1,\dots,x_n)}{C_n(s_n)}\right\}\\
=\min\left\{\lambda_0g_{\theta_0}^{n-1},\lambda_1g_{\theta_1}^{n-1},
g_\theta^{n-1}+\sum_{x_n}U^H_{\theta,n}(s_{n-1}+x_n)C_1(x_n)\frac{C_{n-1}(s_{n-1})}{C_n(s_{n-1}+x_n)}
\right\}\\
\times\frac{h_{n-1}(x_1,\dots,x_{n-1})}{C_{n-1}(s_{n-1})}
=U_{\theta,n-1}^H(s_{n-1})\frac{h_{n-1}(x_1,\dots,x_{n-1})}{C_{n-1}(s_{n-1})}\\
\end{eqnarray*}$\Box$

It is easily seen from the proof that the optimal stopping rule \eqref{11a_3}  is equivalent to
\begin{equation}\label{7n_3}
\psi_n= I_{
\{ \min\{\lambda_0g_{\theta_0}^{n},\lambda_1g_{\theta_1}^{n}\}\leq
g_\theta^n+\mathcal{J}_{n+1}U^H_{\theta,n+1}       \}},\end{equation} 
 and the decision rule  \eqref{11a_4}, to
\begin{equation}\label{7n_4}
\phi_n=I_{\{
\lambda_0g_{\theta_0}^n\leq\lambda_1g_{\theta_1}^n
\}}.
\end{equation}

Now, for any truncation level $H$, we have  optimal tests for the modified Kiefer-Weiss problem in form of \eqref{7n_3}-\eqref{7n_4} (along with their strict-inequality versions), for any fixed $\theta$. Respectively, \eqref{4n_1} acquire the form
\begin{equation}\label{10d_1}\psi_n= I_{
\{ \min\{\lambda_0g_{\theta_0}^{n},\lambda_1g_{\theta_1}^{n}\}\leq
g_\theta^n+\mathcal{J}_{n+1}U_{\theta,n+1}       \}},
\end{equation}
with $U_{\theta,n}  =\lim_{H\to\infty}U_{\theta,n} ^H $ for any natural $n$.

\subsection{Bounds for continuation regions}\label{bounds}
In this part, we find bounds for continuation regions of optimal tests in the modified Kiefer-Weiss problem.

Let us start with the case $\theta\in(\theta_0,\theta_1)$.

It follows from  \eqref{10d_1} that $\psi_n(s)=1$ if $\lambda_0g_{\theta_0}^n(s)< g_{\theta}^n(s)$, and this is equivalent to
\begin{equation}\label{10d_4}
 s> \frac{\log \lambda_0}{\theta-\theta_0}+n\frac{b(\theta)-b(\theta_0)}{\theta-\theta_0}.
\end{equation}
So $\psi_n(s)$ can only be 0 when $s\leq B_n(\lambda_0,\theta_0,\theta)$, where $B_n(\lambda_0,\theta_0,\theta)$ is a maximum integer less  than or equal to the right-hand side of \eqref{10d_4}.

In the same way, $\psi_n(s)=1$ if $\lambda_1g_{\theta_1}^n(s)< g_{\theta}^n(s)$, 
and this  is equivalent to 
\begin{equation}\label{10d_5}
s<\frac{-\log \lambda_1}{\theta_1-\theta}+n\frac{b(\theta_1)-b(\theta)}{\theta_1-\theta},
 \end{equation}
 so $\psi_n(s)$ can only be 0 when $s\geq A_n(\lambda_1,\theta_1,\theta)$, where $A_n(\lambda_1,\theta_1,\theta)$ is a minimum integer 
greater than or equal to the right-hand side of \eqref{10d_5}

Therefore, 
\begin{equation}\label{10d_6}
 \{s:\psi_n(s)<1\}\subset
\{s:A_n(\lambda_1,\theta_1,\theta)\leq s\leq B_n(\lambda_0,\theta_0,\theta)\},
\end{equation}
whenever 
\begin{equation*}\label{10d_7}
 A_n(\lambda_1,\theta_1,\theta)\leq B_n(\lambda_0,\theta_0,\theta).
\end{equation*}
Then, the set of $s$ for which the test continues is finite. 
In addition, there is no such $s$  that $\psi_n(s)<1$ (implying that $\psi_n(s)\equiv 1$) if
\begin{equation*}\label{10d_8}
  A_n(\lambda_1,\theta_1,\theta)>B_n(\lambda_0,\theta_0,\theta),
\end{equation*}
which is equivalent, due to \eqref{10d_4} and \eqref{10d_5}, to
\begin{equation}\label{10d_9}
 \frac{\log \lambda_0}{\theta-\theta_0}+\frac{\log\lambda_1}{\theta_1-\theta}+n\left(\frac{b(\theta)-b(\theta_0)}{\theta-\theta_0}-\frac{b(\theta_1)-b(\theta)}{\theta_1-\theta}\right)<0.
\end{equation}
Let us show that in case $\theta\in(\theta_0,\theta_1)$  the expression in parentheses in \eqref{10d_9} is negative.
Denoting
\begin{equation}\label{10d_10}
G(\theta)=(b(\theta)-b(\theta_0))(\theta_1-\theta)-(b(\theta_1)-b(\theta))(\theta-\theta_0),
\end{equation}
we have $G(\theta_0)=G(\theta_1)=0$, and calculating the second derivative, $G^{\prime\prime}(\theta)
=b^{\prime\prime}(\theta)(\theta_1-\theta_0)$, which is positive because $b^{\prime\prime}(\theta)=\mbox{Var}_\theta X>0$. Thus, $G(\theta)$ is convex, resulting in $G(\theta)<0$ for $\theta\in(\theta_0,\theta_1)$. 
Therefore, \eqref{10d_9} is equivalent to
\begin{equation}\label{12d_1}
n> \left(\frac{\log \lambda_0}{\theta-\theta_0}+\frac{\log(\lambda_1)}{\theta_1-\theta}\right)/\left({\frac{b(\theta_1)-b(\theta)}{\theta_1-\theta}-\frac{b(\theta)-b(\theta_0)}{\theta-\theta_0}}\right).
\end{equation}
Let us denote $H=H(\lambda_0,\lambda_1,\theta_0,\theta_1,\theta)$ the maximum integer less than or equal to  the right-hand side  of \eqref{12d_1}. We know now that  $H$ is the last step when an optimal $\psi_n$ would be allowed to continue, when $\theta\in(\theta_0,\theta_1)$.

There is no bound for $H$ if $\theta\not\in(\theta_0,\theta_1)$.  

If $\theta\geq\theta_1$ then
\begin{equation*}\label{20d_1}\{s:\phi_n(s)<1\}\subset \{s:0\leq s\leq B_n(\lambda_0,\theta_0,\theta)\}
\end{equation*}
for all natural $n$.

At last, if $\theta\leq\theta_0$ then
\begin{equation*}\label{20d_2}\{s:\phi_n(s)<1\}\subset \{s: s\geq A_n(\lambda_1,\theta_1,\theta)\}
\end{equation*}
for all natural $n$.

To conclude this part let us note that the optimal decision rule in  \eqref{7n_4}
can be defined as
$$\phi_n=1-I_{[0,B_n(\lambda_0/\lambda_1,\theta_0,\theta_1)]  }$$
for all natural $n$ (cf. \eqref{10d_4} and the definition of $B_n$).


\subsection{Computational algorithms for designing optimal tests}
\label{alg}

The theoretical basis is formulas \eqref{7n_3} and \eqref{7n_4}.

Let us consider first the case when $\theta\in(\theta_0,\theta_1)$. 

It follows from \eqref{10d_6} that the continuation region is finite (or empty) for any natural $n$. Therefore, the optimal test will be completely characterized by a set of continuation intervals $[a_n,b_n]$, $n=1,\dots, H$, where $H$ is the maximum number of observations the test may take.
 We know from the previous subsection that $H\leq H(\lambda_0,\lambda_1,\theta_0,\theta_1,\theta)$.

Thus, an optimal test can be obtained ``working backward'' from $H=H(\lambda_0,\lambda_1,\theta_0,\theta_1,\theta)$ in the following way.
\begin{enumerate}
\item Set $n=H$.
\item
Define $a_{n-1}$ and $b_{n-1}$, respectively, as a minimum and a maximum  $s\in \mathbb Z^+$, for which 
\begin{equation}\label{18e_1}
g_\theta^{n-1}(s)+\mathcal J_{n}U_{\theta,n}^H(s)<\min\{g_{\theta_0}^{n-1}(s),g_{\theta_1}^{n-1}(s)\}.
\end{equation}

If no such $s$ exist, set $H=n-1$. If $H=1$,  declare ``stop after first step'' state and Stop, else
 go to  Step 1.
\item For $s\in[a_{n-1},b_{n-1}]$, store $v_{n-1}(s)=\min\{g_\theta^{n-1}(s)+\mathcal J_{n}U_{\theta,n}^H(s),g_{\theta_0}^{n-1}(s),g_{\theta_1}^{n-1}(s)\}$. Take into account, for future use, that
$$U_{\theta,n-1}^H(s)=\begin{cases}v_{n-1}(s), &\mbox{if}\, s\in [a_{n-1},b_{n-1}]\cr
\min\{g_{\theta_0}^{n-1}(s),g_{\theta_1}^{n-1}(s)\},&\mbox{ otherwise}.\end{cases}
$$
 \item If $n=2$, then Stop, else set $n=n-1$, and go to Step 2.
\end{enumerate}

``Stop after first step'' as a result of this algorithm means that the optimum test has to stop after the first observation is taken (thus it is not sequential). Usually, this means that $\lambda_0$ and/or $\lambda_1$ are too small to produce a meaningful sequential test. An optimal test can be defined in this case as $\psi_1\equiv 1$ and $\phi_1=1-I_{  [0,B_1(\lambda_0/\lambda_1,\theta_0,\theta_1)]}(s)$.

If the algorithm does not terminate in ``stopping after first step'', we have an optimal test with continuation intervals $[a_n,b_n]$, $n=1,2, \dots, H-1$, with a maximum number of observations equal to  $H$.

If $\theta\not\in(\theta_0,\theta_1)$, the above algorithm can be used, after some modifications, for obtaining optimal tests in the class of {\em truncated} tests $\mathscr M_\theta^H$. The modifications concern the way the ``backward induction" in Step 2 is performed, because there is only one bound for the continuation region in any one of the cases seen in Subsection \ref{bounds}, when $\theta\not\in(\theta_0,\theta_1)$. For example, if $\theta>\theta_1$, we  seek for $s$ satisfying \eqref{18e_1} downwards,
starting from $B_{n-1}(\lambda_0,\theta_0,\theta)$, then, starting from $s=0$, upwards, obtaining, respectively, $b_{n-1}$ and $a_{n-1}$ (if any).

If $\theta=\theta_1$, we can narrow the search region in Step 2, because it is known \citep[see][]{NovikovPopoca} that the continuation region (if not empty) is always an interval when sampling from an exponential family, so, after $b_{n-1}$ has been found, we may want to keep seaching for $a_{n-1}$ downwards.

There is a very similar situation in the case
$\theta=\theta_0$, where we search for $a_{n-1}$ upwards, starting from $A_{n-1}(\lambda_1,\theta_1,\theta_0)$,  and then keep searching upwards, until $b_{n-1}$ is found.

The only remaining case $\theta<\theta_0$ is the hardest one, because there is no upper bound for the continuation region, unless the distribution of $S_{n-1}$ is bounded (as in the binomial case).
So, in our computer implementation, we search for $a_{n-1}$ upwards, starting from $A_{n-1}(\lambda_1,\theta_1,\theta_0)$,  and,  after $a_{n-1}$ has been found, keep searching upwards for $b_{n-1}$
 indefinitely. When $b_{n-1}$ is found, it will  correspond to the optimal test unless there is 
an $s>b_{n-1}$ satisfying \eqref{18e_1}. Unfortunately, there is no theoretical result which could guarantee that the continuation region is an interval in this case. Anyway, we implement the algorithm as described above, so one has to take into account that the test obtained in the modified Kiefer-Weiss problem in case $\theta<\theta_0$ can be suboptimal. The exact algorithm we implemented for the Bernoulli case in \cite{novikov2021comp}, in particular cases we experimented with, gives the same results.

 As shown in \cite{Hawix}, the optimal tests for modified Kiefer-Weiss problem with $\theta>\theta_1$ or $\theta<\theta_0$ may not stop with probability one under H$_0$ and/or H$_1$,  so they are not of particular importance for applications. Our finite-horizon implementation in this Section may be useful in these situations, because it terminates wth probability one under each one of the hypotheses and gets close to the optimal modified Kiefer-Weiss test, as $H\to\infty$. Nevertheless, high values of the average sample number should be expected where they are infinite in the infinite-horizon case. Instead, one may want to include the average sample number, whose large value causes the problem, as an additional criterion for the Lagrangian minimization, with some small weight. For example, if one wants to minimize $N(\theta;\psi)$ with $\theta<\theta_0$, the Lagrangian function may include, additionally, a term $cN(\theta_1;\psi)$ with some small $c$. The only change to the algorithm above would be using
\begin{equation*}\label{21e_1}
g_\theta^{n-1}(s)+cg_{\theta_1}^{n-1}(s)+\mathcal J_{n}U_{\theta,n}^H(s)<\min\{g_{\theta_0}^{n-1}(s),g_{\theta_1}^{n-1}(s)\}.
\end{equation*}
in place of \eqref{18e_1}. Probably, this could be helpful in the applied problem mentioned  by \cite{Hawix}, where the modified Kiefer-Weiss problem is shown to be largely useless.
In no way, this will be a Kiefer-Weiss problem any more, so we do not go into detail.
Interested readers can easily  modify our code on their own to obtain a meaningful test for this problem. The rest of algorithms for performance evaluation we develop below will be applicable without any change.
%
%
\subsection{Operating characteristic, average sample number and related formulas}\label{opch}

In this Section, we obtain formulas for calculating error probabilities, average sample number amd some  related probabilities, for  truncated sequential tests.

Let $\langle\psi,\phi\rangle\in\mathscr S^H$. This test will be held fixed within this subsection, so   we will suppress it in  the notation.

Let \begin{equation}\label{25a_1}a_\theta^H(s)=1-\phi_H(s),\; s\in\mathbb Z^+,\end{equation} and, recursively over $n=H-1,H-2,\dots,1$, 
\begin{equation*}\label{25a_2}a_\theta^{n}(s)=\psi_{n}(s)(1-\phi_n(s))+(1-\psi_{n}(s))E_{\theta}a_\theta^{n+1}(s+X_{n+1}),\;s\in\mathbb Z^+.\end{equation*}

Let us denote $D_n^H=D_n^H(\psi,\phi)$ the event meaning 
 ``H$_0$ is accepted at or after step $n$ as a result of applying the test $\langle \psi,\phi\rangle$''. In particular, the operating characteristic curve is $OC_\theta(\psi,\phi)=P_\theta(D_1^H(\psi,\phi))$. 

 Let $S_n=\sum_{i=1}^n X_i$ for any natural $n$, and $S_0=0$.

\begin{proposition}\label{P2} 
For any  $1\leq n\leq H$
\begin{equation}\label{22e_1} a_\theta^n(S_n)=P_\theta({D_n^H}| X_1,\dots, X_n),\end{equation}  and 
$$
OC_\theta(\psi,\phi)=E_\theta a_\theta^1(X_1).
$$
\end{proposition}

{\bf Proof.} By induction over $n=H,H-1,\dots,1$.

For $n=H$, it follows from \eqref{25a_1} that $
a_\theta^H(S_H)=1-\phi_H(S_H)=P_\theta \{D_H^H| X_1,\dots,X_H\}$
(this latter equality follows from the definition of the decision function $\phi$).

Let us suppose now that \eqref{22e_1} holds for some $2\leq n\leq H$.

Then \begin{eqnarray*}
a_\theta^{n-1}(S_{n-1}) &=&
 \psi_{n-1}(S_{n-1})(1-\phi_{n-1}(S_{n-1}))\\
&&+(1-\psi_{n-1}(S_{n-1}))E_\theta\{a_{\theta}^{n}(S_{n-1}+X_n)| X_1,\dots,X_{n-1}\}\\
&=&\psi_{n-1}(S_{n-1})(1-\phi_{n-1}(S_{n-1}))\\
&&+ (1-\psi_{n-1}(S_{n-1}))E_\theta\{P_\theta\{D_n^H|X_1,\dots,X_n\}| X_1,\dots,X_{n-1}\}\\
&=&\psi_{n-1}(S_{n-1})(1-\phi_{n-1}(S_{n-1}))\\
&&+(1-\psi_{n-1}(S_{n-1}))E_\theta\{I_{D_n^H}| X_1,\dots,X_{n-1}\}\\
&=&E_\theta\{\psi_{n-1}(S_{n-1})(1-\phi_{n-1}(S_{n-1}))
+(1-\psi_{n-1}(S_{n-1}))I_{D_n^H}| X_1,\dots,X_{n-1}\}\\
&=&P_\theta\{D_{n-1}^H| X_1,\dots, X_{n-1}\}
  \end{eqnarray*}
$\Box$

In a similar way, let for any $k<H$ \begin{equation}\label{23e_1}
                       b_{\theta,k}^k(s)=(1-\psi_k(s)),\; s\in\mathbb Z^+,
                      \end{equation}
 and, recursively over $n=k-1,k-2,\dots,1$, 
\begin{equation}\label{23e_2}
b_{\theta,n}^{k}(s)=E_\theta b_{\theta,n+1}^{ k     }(s+X_{n+1})(1-\psi_n(s)),\,s\in \mathbb Z^+
\end{equation}

\begin{proposition}\label{P3}For all $n\leq k<H$
\begin{equation}\label{23e_3}
b_{\theta,n}^k(S_n)=E_\theta\{(1-\psi_n)\dots(1-\psi_k) |X_1,\dots,X_n\}.
\end{equation}
In particular,
 $P_\theta(\tau_\psi>k)=E_\theta b_{\theta,1}^k(X_1)$.
\end{proposition}
{\bf Proof.} By induction over $n=k,k-1,\dots,1$.

For $n=k$, it follows from \eqref{23e_1} that $
b_\theta^k(S_k)=1-\psi_k(S_k)=E_\theta \{1-\psi_k(S_k)| X_1,\dots,X_k\}$.

Let us suppose now that \eqref{23e_3} holds for some $2\leq n\leq k$.

Then \begin{eqnarray*}
b_{\theta,{n-1}}^k(S_{n-1}) &=&
(1-\psi_{n-1}(S_{n-1}))E_\theta\{b_{\theta,n}^{k}(S_{n-1}+X_n)| X_1,\dots,X_{n-1}\}\\
&=
& (1-\psi_{n-1}(S_{n-1}))E_\theta\{E_\theta\{(1-\psi_n)\dots(1-\psi_k))|X_1,\dots,X_n\}| X_1,\dots,X_{n-1}\}\\
&=&(1-\psi_{n-1}(S_{n-1}))E_\theta\{(1-\psi_n)\dots(1-\psi_k)| X_1,\dots,X_{n-1}\}\\
&=&E_\theta\{(1-\psi_{n-1})\dots(1-\psi_k)| X_1,\dots, X_{n-1}\}
  \end{eqnarray*}
$\Box$

The average sample number can now  be calculated  as $N(\theta;\psi)=
\sum_{k=1}^HP_\theta(\tau_\psi\geq k)$.

A more direct way is to  incorporate the calculation of this sum  into \eqref{23e_2}. In fact, we can apply the algorithm to any stopping rule $\psi=(\psi_1,\psi_2,\dots)$ by truncating it at level $H$, that is, defining $\psi^H=(\psi_1,\dots, \psi_H\equiv 1,\dots)$. 

\begin{corollary}\label{C1}

Let  \begin{equation*}\label{23e_4}
                       d_{\theta,H-1}^H(s)=(1-\psi_{H-1}(s)),\; s\in\mathbb Z^+,
                      \end{equation*}
 and, recursively over $n=H-2,H-3,\dots,1$, 
\begin{equation*}\label{23e_5}
d_{\theta,n}^{H}(s)=(1+E_\theta d_{\theta,n+1}^{  H    }(s+X_{n+1}))(1-\psi_n(s)),\,s\in \mathbb Z^+.
\end{equation*}

Then $N(\theta;\psi^H)=1+E_\theta d_{\theta,1}^H(X_1)$.
 
\end{corollary}
We  consider meaningful for applications only the tests for which $N(\theta;\psi^H)\to N(\theta;\psi)$, as $H\to\infty$. Thus, Corollary \ref{C1} provides a way for numerical evaluation of the average sample number $N(\theta;\psi)$ as $\lim_{H\to\infty}(1+E_\theta d_{\theta,1}^H(X_1))$.
In particular, we use this fact for calculating the average sample size of the SPRT and of the optimal tests in the modified Kiefer-Weiss problems when $\theta\not \in (\theta_0,\theta_1)$.
\subsection{Numerical results}
The main goal of this part is to illustrate the use of the developed algorithms on concrete examples of discrete exponential  families of distributions,  to show that the obtained sequential tests in any one of the examples provide numerical solutions to the Kiefer-Weiss problem, and to analyze the efficiency of the obtained tests with respect to the classical sequential probability ratio tests and the fixed-sample-size tests, provided these have the same level of error probabilities.

We use a series of concrete examples 
of hypotheses tests for binomial, Poisson and geometric distributions.
We test the following hypotheses: $\theta_0=0.05$ vs. $\theta_1=0.08$ for the binomial distribution corresponding to  $n=3$ Bernoulli trials with success probability $\theta$;  $\theta_0=0.5$ vs. $\theta_1=0.7$ for the Poisson distribution with mean $\theta$; and $\theta_0=1$ vs. $\theta_1=2$ for the geometric distribution with mean $\theta$. 
For each pair, we employ a range of error probabilities widely used in the  practice: $\alpha(=\beta)=$ 0.1, 0.05, 0.025, 0.01, 0.005, 0.001 and 0.0005.

 In addition, to see the effect of asymmetric error probabilities we made the same evaluations for  $\alpha=0.1$ and $\beta=0.0005$ for each pair of the hypotheses. 

For each combination of $\theta_0,\theta_1$ and $\alpha$ and $\beta$ we ran the computer code corresponding to the implementation of the method of Section 2 in \cite{novikov2021comp}(using Option 1 with the bound $H$ defined by the right-hand side  of \eqref{12d_1}). For the solution of the corresponding modified Kiefer-Weiss problem, we use the algorithms of 
Subsections \ref{alg} and \ref{opch}.

To comply with the restrictions on the error probabilities, we seek for the Lagrange multipliers $\lambda_0$ and $\lambda_1$ that provide the best approximation to the nominal values of $\alpha$ and $\beta$.
We use the numerical minimization, over $\lambda_0$ and $\lambda_1$ searching for a minimum value of \begin{equation}\label{14e_1}\max\{|\alpha(\psi^*,\phi^*)-\alpha |/\alpha, |\beta(\psi^*,\phi^*)-\beta |/\beta\}.\end{equation}
We employed the {\tt fminsearch} function of the {\tt neldermead} R package for the gradient-free Nelder-Mead method \citep[see][]{neldermead}.
A reasonably good formula to start the numerical minimization of \eqref{14e_1} is 
$
\lambda_0=\kappa_0 /\alpha, \quad \lambda_1=\kappa_1 /\beta.
$
where $\kappa_0$ and
$\kappa_1$ are empirical coefficients in the range of 10 to 60 depending on the model.

Due to the discrete nature of the probabilities involved in the evaluation there is  no guarantee, generally speaking, that the minimum of \eqref{14e_1} can  be 0.
Nevertheless, in
all the evaluated cases the real and the nominal error probability are approximately within 0.002 of relative distance \eqref{14e_1} to each other, for the  range of $\alpha$ and $\beta$ evaluated.

The respective numerical results are presented in Tables 1  -- 3. For each test, the table contains the corresponding values of $\theta,\lambda_0,\lambda_1$,  its average sample number $N(\theta;\psi^*)$   and its corresponding $\Delta$ (see \eqref{20m_1}), as well as  the average sample number under the two hypotheses, $N(\theta_0;\psi^*)$ and $N(\theta_1;\psi^*)$, and the 0.99-quantile Q$_{.99}(\psi^*) $ of the distribution of the sample number, under $\theta$. We also  present the maximum sample number (denoted $H$ in the table) the test actually takes.

In the second part of each table, there are the calculated characteristics of the corresponding SPRT with the closest values of $\alpha$ and $\beta$ to the nominal ones. The same numerical minimization procedure as above has been employed, this time over the continuation bounds $A$ and $B$ of the SPRT.
We place in the table the values of the average sample number of  Wald's SPRT $N(\theta;W)$,
along with  the average sample number under both hypotheses, $N(\theta_0;W)$ and $N(\theta_1;W)$, and the 0.99-quantile Q$_{.99}(W)$ of the  distribution of the sample number,  calculated at $\theta$. All the characteristics are calculated  using the exact  formulas in Subsection \ref{opch},
with appropriately large horizon $H$. 
$\log (A) $ and $\log (B)$ are the endpoints of the continuation interval of the corresponding SPRT.

At last, FSS is the minimum value  of the sample number required by the optimal  fixed-sample-size test with error probabilities  $\alpha$ and $\beta$. For given $\alpha$ and $\beta$, FSS is calculated as $$n^*+\frac{\beta(n^*)-\beta}{\beta(n^*)-\beta(n^*+1)},$$ 
where $n^*=n^*(\alpha,\beta)$ is a maximum integer $n$ for which the type II-error probability $\beta(n)$  of  the most powerful level-$\alpha$  (Neyman-Pearson) test is greater than or equal to $\beta$.

In the last part of each table, there are calculated values of efficiency of each test with respect to the FSS tests. The efficiency is calculated as the ratio of FSS to other characteristics of the respective test:  $R(\psi^*)={FSS}/{N(\theta;\psi^*)}$, $R_0(\psi^*)={FSS}/{N(\theta_0;\psi^*)}$, $R_1(\psi^*)={FSS}/{N(\theta_1;\psi^*)}$ and $QR(\psi^*)={FSS}/{Q_{.99}(\psi^*)}$ for the optimal Kiefer-Weiss test and $R(W)={FSS}/{N(\theta;W)}$, $R_0(W)={FSS}/{N(\theta_0;W)}$, $R_1(W)={FSS}/{N(\theta_1;W)}$ and $QR(W)={FSS}/{Q_{.99}(W)}$ for the Wald's SPRT. For example, $R_0(\psi^*)=2$ means that the optimal Kiefer-Weiss test takes   2 times fewer observations, on the average, under $H_0$, than the corresponding fixed-sample-size test.

\begin{table}[!p]
\begin{tabular}{c|cccccccc}
$\alpha=\beta\to$&0.1&0.05&0.025&0.01&0.005&0.001&0.0005\\
\hline
$\lambda_0$&305.94&691.65&1495.23&3993.13&8275.87&43707.77&88888.07\\
$\lambda_1$&326.39&737.05&1596.43&4262.75&8815.25&46564.03&94690.75\\
$\theta$&0.58464&0.58794&0.58953&0.59072&0.59130&0.59206&0.59225\\
 $H$&353&442&533&649&745&935&1024\\
$ N(\theta;\psi^*)$&67.93&114.80&166.46&239.77&297.96&439.60&502.73\\
$ N(\theta_0;\psi^*)$&57.06&87.92&117.97&156.92&185.92&252.90&281.79\\
$ N(\theta_1;\psi^*)$&51.66&79.55&106.80&141.98&168.15&228.59&254.63\\
$Q_{.99}(\psi^*)$&165&247&331&440&522&712&794\\
$\Delta$&1E-06&2E-06&2E-05&9E-06&3E-06&8E-06&4E-05\\
\hline
log(A)&-0.916&-1.240&-1.553&-1.957&-2.260&-2.961&-3.262\\
log(B)&0.868&1.191&1.504&1.908&2.212&2.912&3.214\\
$ N(\theta;W)$&72.28&129.16&199.74&313.98&416.52&708.62&857.74\\
$ N(\theta_0;W)$&55.55&83.53&109.64&141.86&165.01&217.05&239.08\\
$ N(\theta_1;W)$&50.06&75.11&98.48&127.26&148.04&194.56&214.22\\
$Q_{.99}(W)$&281&504&780&1229&1632&2779&3364\\
\hline
FSS&98.07&161.05&229.01&322.43&395.11&568.75&644.84\\
\hline
$R(\psi^*)$&1.44&1.40&1.38&1.34&1.33&1.29&1.28\\
$R_0(\psi^*)$&1.72&1.83&1.94&2.05&2.13&2.25&2.29\\
$R_1(\psi^*)$&1.90&2.02&2.14&2.27&2.35&2.49&2.53\\
$QR(\psi^*)$&0.59&0.65&0.69&0.73&0.76&0.80&0.81\\
$R(W)$&1.36&1.25&1.15&1.03&0.95&0.80&0.75\\
$R_0(W)$&1.77&1.93&2.09&2.27&2.39&2.62&2.70\\
$R_1(W)$&1.96&2.14&2.33&2.53&2.67&2.92&3.01\\
$QR(W)$&0.35&0.32&0.29&0.26&0.24&0.20&0.19\\
\end{tabular}
\caption{  Poisson distribution with mean $\theta_0=0.5$ vs. $\theta_1=0.7$}
\end{table}
\begin{table}[!p]
\begin{tabular}{c|cccccccc}
$\alpha=\beta\to$& 0.1&0.05&0.025&0.01&0.005&0.001&0.0005\\
\hline
$\lambda_0$ &69.00&154.09&333.06&893.28&1848.87&9762.77&19882.77\\
$\lambda_1$& 84.38&189.88&408.32&1092.24&2258.64&11940.38&24318.97\\
$\theta$&1.27794&1.31841&1.33953&1.35556&1.36336&1.37428&1.37741\\
 $H$&74&96&118&145&167&213&233\\
$ N(\theta;\psi^*)$&17.08&28.37&40.81&58.44&72.52&106.58&121.76\\
$N(\theta_0;\psi^*)$&
15.43&23.60&31.62&41.92&49.66&67.51&75.212\\
$ N(\theta_1;\psi^*)$&
11.63&17.53&23.35&30.75&36.32&49.11&54.57\\
$Q_{.99}(\psi^*)$&41&60&80&106&126&172&192\\
$\Delta$& 3E-07&1E-08&1E-06&9E-09&8E-08&-2E-07&-5E-08\\
\hline
log $A$&-0.8920&-1.2098&-1.5170&-1.9213&-2.2260&-2.9269&-3.2278\\
log $B$&0.7318&1.0452&1.3615&1.7649&2.0668&2.7692&3.0698\\
$ N(\theta;W)$&18.24&31.59&48.32&75.44&99.78&168.99&204.24\\
$ N(\theta_0;W)$&15.30&22.62&29.64&38.30&44.53&58.60&64.53\\
$ N(\theta_1;W)$&11.42&16.61&21.48&27.51&31.87&41.60&45.71\\
$Q_{.99}(W)$&67&119&184&290&384&654&791\\
\hline
FSS&23.83&39.00&55.29&77.74&95.21&136.89&155.21\\
\hline
$R(\psi^*)$& 1.40&1.37&1.35&1.33&1.31&1.28&1.27\\
$R_0(\psi^*)$&1.54&1.65&1.75&1.85&1.92&2.03&2.06\\
$R_1(\psi^*)$&2.05&2.22&2.37&2.53&2.62&2.79&2.84\\
$QR(\psi^*)$&0.58&0.65&0.69&0.73&0.76&0.80&0.81\\
$R(W)$&1.31&1.23&1.14&1.03&0.95&0.81&0.76\\
$R_0(W)$&1.56&1.72&1.87&2.03&2.14&2.34&2.41\\
$R_1(W)$&2.09&2.35&2.57&2.83&2.99&3.29&3.40\\
$QR(W)$&0.36&0.33&0.30&0.27&0.25&0.21&0.20\vspace{2mm}\\
\end{tabular}
\caption{Geometric distribution with mean $\theta_0=1$ vs. $\theta_1=2$}
\end{table}
\begin{table}[!p]
\begin{tabular}{c|cccccccc}
$\alpha=\beta\to$& 0.1&0.05&0.025&0.01&0.005&0.001&0.0005\\
\hline

$\lambda_0$&450.00&1020.19&2201.68&5880.48&12177.81&64446.61&131326.7\\
$\lambda_1$&489.75&1110.18&2393.14&6398.75&13250.02&70018.20&142274.9\\
$\theta$&0.06193&0.06263&0.06296&0.06316&0.06328&0.06345&0.06350\\
$H$&484&630&771&943&1068&1355&1480\\
$N(\theta;\psi^*)$&101.13&171.00&247.90&357.06&443.72&654.74&748.81\\
$N(\theta_0;\psi^*)$&86.28&133.00&178.58&237.31&281.04&382.50&426.12\\
$N(\theta_1;\psi^*)$&75.46&116.41&156.41&208.08&246.49&335.21&373.45\\
$Q_{.99}(\psi^*)$&247&369&493&654&778&1060&1184\\
$\Delta$&-3E-07&-1E-06&2E-04&2E-04&4E-04&-7E-08&5E-05\\
\hline
log(A)&-2.1517&-2.8990&-3.6164&-4.5494&-5.2469&-6.8607&-7.5540\\
log(B)&2.0034&2.7527&3.4713&4.4033&5.1012&6.7147&7.4086\\
$N(\theta;W)$&107.24&192.16&296.92&466.76&618.96&1053.14&1275.06\\
$N(\theta_0;W)$&83.91&126.52&166.12&215.01&250.13&329.02&362.47\\
$N(\theta_0;W)$&73.07&109.94&144.16&186.36&216.82&285.00&313.86\\
$Q_{.99}(W)$&414&748&1159&1826&2423&4124&4997\\
\hline
FSS&146.62&240.94&341.25&480.98&589.32&847.40&961.23\\
\hline
$R(\psi^*)$&1.45&1.41&1.38&1.35&1.33&1.29&1.28\\
$R_0(\psi^*)$&1.70&1.81&1.91&2.03&2.10&2.22&2.26\\
$R_1(\psi^*)$&1.94&2.07&2.18&2.31&2.39&2.53&2.57\\
$QR(\psi^*)$&0.59&0.65&0.69&0.74&0.76&0.80&0.81\\
$R(W)$&1.37&1.25&1.15&1.03&0.95&0.80&0.75\\
$R_0(W)$&1.75&1.90&2.05&2.24&2.36&2.58&2.65\\
$R_1(W)$&2.01&2.19&2.37&2.58&2.72&2.97&3.06\\
$QR(W)$&0.35&0.32&0.29&0.26&0.24&0.21&0.19\\

\end{tabular}
\caption{Binomial distribution with  $\theta_0=0.05$ vs. $\theta_1=0.08$ and $n=3$}
\end{table}
\begin{table}[!p]
\begin{tabular}{c|ccc}
&Binomial&Poisson&Geometric\\
& ($n=3$)\\
\hline
$\theta_0$/$\theta_1$&0.05/0.08&0.5/0.7&1/2\\
$\lambda_0$&948.57&640.92&150.43\\
$\lambda_1$&91786.79&60360.17&15823.13\\
$\theta$&0.05551&0.53918&1.12992\\
$H$&1172&806&200\\
$N(\theta;\psi^*)$&350.27&232.95&60.47\\
$N(\theta_0;\psi^*)$&320.30&211.46&56.82\\
$N(\theta_1;\psi^*)$&116.76&79.88&17.36\\
$Q_{.99}(\psi ^*)$&695&464&118\\
$\Delta$&-8E-05&6E-06&-5E-08\\
\hline
log(A)&-7.4481&-3.2167&-3.1829\\
log(B)&2.1115&0.9132&0.7673\\
$N(\theta;W)$&392.78&263.20&65.89\\
$N(\theta_0;W)$&311.04&205.19&55.50\\
$N(\theta_1;W)$&95.35&65.17&14.54\\
$Q_{.99}(W)$&1293&879&200\\
\hline
$FSS$&478.61&319.13&81.10\\
\hline
$R(\psi^*)$&1.37&1.37&1.34\\
$R_0(\psi^*)$&1.49&1.51&1.43\\
$R_1(\psi^*)$&4.10&4.00&4.67\\
$QR(\psi^*)$&0.69&0.69&0.69\\
$R(W)$&1.22&1.21&1.23\\
$R_0(W)$&1.54&1.56&1.46\\
$R_1(W)$&5.02&4.90&5.58\\
$QR(W)$&0.37&0.36&0.41\\
\end{tabular}
\caption{Optimal tests for $\alpha=0.1$, $\beta=0.0005$}
\end{table}

\section{Discussion and conclusions}
\subsection{Analysis of numerical results}
Tables 1 -- 4 are convenient for analyzing the relative  efficiency of the optimal tests with respect to the fixed-sample-size test. 

It is clearly seen that all the three efficiencies ($R(\psi ^*)$, $R_0(\psi ^*)$ ,$R_1(\psi ^*)$  do not  vary much in the whole range of $\alpha$ and $\beta$ computed. 

The lowest value of $R(\psi ^*)$  is about 1.3 and is attained at the minimum values of $\alpha =\beta$ considered. There is a clear tendency of the relative efficiency decreasing with $\alpha=\beta\to 0$. As a reference, one can bear in mind that the relative efficiency $R(\psi^*)$ can not drop below 1, by its definition. 

Both $R_0(\psi^*)$ and $R_1(\psi^*)$ slightly vary between approx. 1.8 and 2.8 for  $\alpha=\beta$.

It is clearly seen from Table 4 that $R(\psi^*)$  maintains its level at approx. 1.4, even for very asymmetric $\alpha$ and $\beta$.

The relative efficiency under H$_1$, $R_1$, tends to have higher values (up to approx. 4.0 to 4.5) for very asymmetric case  of  $\alpha=0.1$ and $\beta=0.0005$, while $R_0(\psi^*)$  stays at approx. 1.5, which is slightly lower than in the case of equal $\alpha$ and $\beta$.
 
The relative efficiency $QR(\psi^*)$ based on the 0.99-quantile of the optimal Kiefer-Weiss test behaves quite well maintaining  the approximate level of 0.6 to 0.8, for all levels of $\alpha, \beta$ computed.

The relative efficiency of the SPRT based on the average sample number evaluated at $\theta$ drops to approx. 0.7 -- 0.8 for lower levels of $\alpha$ and $\beta$, which  is still comparable to the efficiency of approx. 1.3 of the optimal Kiefer-Weiss test. But SPRT shows a very low efficiency $QR(W)$ based on the 0.99-quantile of the sample number distribution, under $\theta$, which is as low as approx. 0.3 -- 0.2, meaning the 0.99-quantile can reach a 3 -- 5 times higher level than the fixed sample size.

In general, it seems remarkable that the pattern of efficiency is largely the same both between the three models we consider in this paper, and between these ones and the Bernoulli model in \cite{novikov2021comp}. 

Comparing the relative efficiency numbers of $R(\psi^*)$, for the respective $\alpha=\beta$ level between the four models, we observe almost identical results.
$R_0(\psi^*)$  and $R_1(\psi^*)$ show very similar behaviour. 

The same happens with the efficiencies calculated for the  SPRT (comparing $R(W)$ or $R_0(W)$, or $R_1(W)$ between the models). Even quantile-based efficiencies $QR(\psi^*)$ are practically identical  between the models, and  the same is valid for $QR(W)$.

\subsection{Further work}

The most immediate work to be done is to develop the computer algorhithms for solving the Kiefer-Weiss problem for sampling from continuous exponential families of distributions. For normal distribution and $\alpha=\beta$ there are various known numerical results in the literature \citep[see, for example,][]{Lorden76}. 

Another important aspect is the efficiency of Lorden's 2-SPRT with respect to the optimal Kiefer-Weiss test. The numerical results of \cite{Lorden76} show  an  excellent performance of the 2-SPRT in the symmetric normal case. Our method should not only permit to investigate 
non-symmetric normal case, but also the performance of the 2-SPRT for other exponential families of distributions. In particular, assessing non-asymptotic rate of approximation for  the asymptotically optimal 2-SPRT constructed by \cite{Huffman1983} could be of interest for applications like \cite{mulekar}.  

\section*{Acknowledgments}
A. Novikov gratefully acknowledges a partial support of SNI by CONACyT (Mexico) for this work. F. Farkshatov thanks CONACyT (México) for scholarship for his doctoral studies.

\end{document}